\documentclass[twocolumn,floatfix,showpacs,pra]{revtex4}
\usepackage{color}
\usepackage{mathbbol}              
\usepackage{graphics,graphicx,epsfig,ulem,amsmath,eucal,bm}
\def\bra#1{\mathinner{\langle{#1}|}}
\def\ket#1{\mathinner{|{#1}\rangle}}

\begin{document}
\title{Sub-natural linewidths in excited-state spectroscopy}
\author{M. Tanasittikosol, C. Carr, C. S. Adams and K. J. Weatherill}
\affiliation{Department of Physics, Durham University,
Rochester Building, South Road, Durham DH1 3LE, UK}


\begin{abstract}
We investigate, theoretically and experimentally, absorption on an excited-state atomic transition in a thermal vapor where the lower state is coherently pumped. We find that the transition linewidth can be sub-natural, i.e. less than the combined linewidth of the lower and upper state. For the specific case of the 6P$_{3/2}$ $\rightarrow$ 7S$_{1/2}$ transition in room temperature cesium vapor, we measure a minimum linewidth of 6.6~MHz compared with the natural width of 8.5~MHz. Using perturbation techniques, an expression for the complex susceptibility is obtained which provides excellent agreement with the measured spectra.
\end{abstract}
\maketitle


\section{Introduction}
Spectroscopy of excited state transitions is of growing interest for a variety of applications including the search for stable frequency references \cite{bret93,abel09}, state lifetime measurement \cite{shen08}, optical filtering \cite{bill95}, frequency up-conversion \cite{meij06}, multi-photon laser cooling \cite{wu09}, as well as Rydberg gases \cite{weat08,kubl11} and their application to electro-optics \cite{moha08,baso10,taus10} and non-linear optics \cite{prit10}.
Excited state spectroscopy can be achieved without significant transfer of population out of the ground state using electromagnetically induced transparency (EIT) \cite{eit_review} in the ladder configuration \cite{bana95}. In conventional EIT, the excited state transition is driven by a strong coupling laser creating a transparency window which is then detected using a weak probe on the ground state transition. In thermal vapors, ladder EIT is only possible when the lower transition is probed \cite{anis11} and the probe wavelength is greater than the coupling wavelength \cite{shep96}. Alternatively, on strong transitions such as the infra-red transitions from excited states in alkali atoms, one can probe directly on the excited state transition and detect absorption or fluorescence \cite{fox93}.

%
\begin{figure}[h]
\begin{center}
\includegraphics[width=8.2cm]{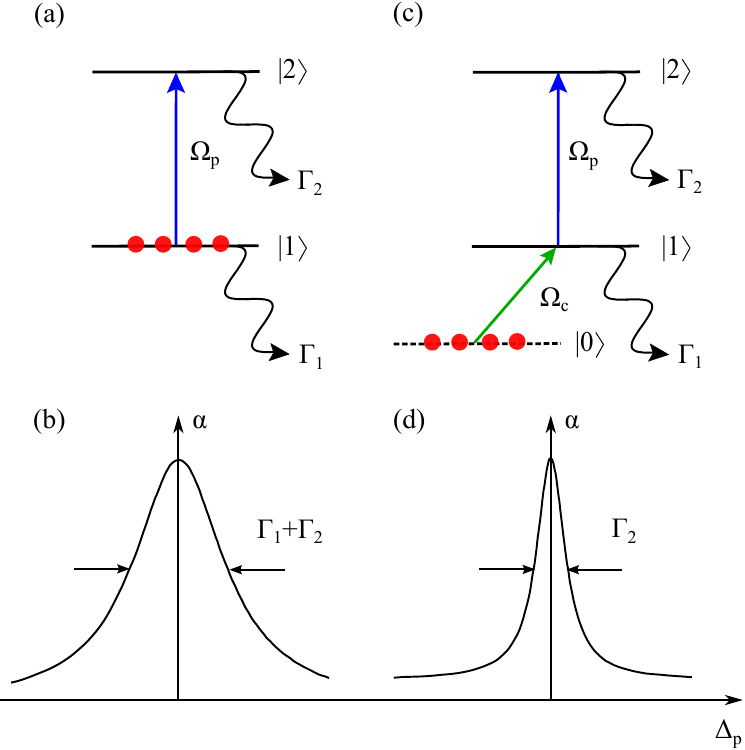}
\caption[]{(Color online) (a) Schematic of the energy levels of an atom whose lower state $\ket{1}$ is incoherently populated. The natural decay of states $|1\rangle$ and $|2\rangle$ are $\Gamma_1$ and $\Gamma_2$, respectively. The absorption lineshape is probed by the probe field of Rabi frequency $\Omega_{\rm p}$. (b) The FWHM of the lineshape is $\Gamma_1+\Gamma_2$. (c) Schematic of the energy levels of a transition whose lower state $\ket{1}$ is coherently populated by the coupling field, of Rabi frequency $\Omega_{\rm c}$, from a state $\ket{0}$. In this case the FWHM of the absorption lineshape is $\Gamma_2$ when $\Omega_{\rm c}$ is small, as shown in (d). \label{fig1}}
\end{center}
\end{figure}

In this work we develop the theory for the complex susceptibility of excited state transitions and compare the results to experimental observations. We consider a transition between two states, $|1\rangle$ and $|2\rangle$ shown in Fig.~1~(a). Neglecting the effects of Doppler broadening, it is expected that the lineshape of the absorption is a Lorentzian whose FWHM is given by the sum of the natural linewidths of states $|1\rangle$ and $|2\rangle$, $\Gamma_1+\Gamma_2$, as shown in Fig. 1(b). We show that if the population of state $|1\rangle$ is coherently pumped from another state with a weak coupling field, Fig.~1(c), the absorption lineshape remains a Lorentzian but its FWHM is solely determined by the natural linewidth of the upper state, $\Gamma_2$ as shown in Fig.~1(d) because the lower state is effectively stable.

The paper is organized as follows. In Section 2 we develop the equations of motion for the system and derive an expression for the susceptibility. In Section 3 we extend the analysis to include the effects of Doppler broadening. In Section 4 we compare our derived expressions to experimental data in the limit of weak pumping. In Section 5 we compare theory and experiment in the regime of strong coupling before concluding in Section 6.


\section{Excited state transition with coherent pumping of lower state}\label{sec:sec2}
\subsection{Equations of motion and the steady state solutions}
Consider a transition between 2 states as shown in Fig.~\ref{fig1}(c), a lower state, $\ket{1}$, and an upper state, $\ket{2}$, with the associated eigenenergies of $\hbar\omega_1$ and $\hbar\omega_2$, respectively. Initially, states $|1\rangle$ and $|2\rangle$ are not populated. To populate state $|1\rangle$, the system is coherently pumped by the resonant coupling field with Rabi frequency of $\Omega_{\rm c}$ from the stable eigenstate $\ket{0}$ whose eigenenergy is $\hbar\omega_0$ and $\omega_0<\omega_1<\omega_2$. The transmission (or absorption) lineshape of the transition is probed by scanning the frequency of the probe field whose Rabi frequency is $\Omega_{\rm p}$. Applying the rotating-wave approximation and the slowly-varying variables transformation, the Hamiltonian of the system is given by
${\cal H}={\cal H}_0+{\cal H}_{\rm I}$, where
\begin{subequations}
\begin{flalign}
{\cal H}_0&=-\hbar\Delta_{\rm c}\ket{1}\bra{1}-\hbar\Delta_{\rm R}\ket{2}\bra{2}~,\\
{\cal H}_{\rm I}&=\frac{\hbar\Omega_{\rm p}}{2}\ket{2}\bra{1}+\frac{\hbar\Omega_{\rm c}}{2}\ket{1}\bra{0}+{\rm h.c.}~,
\end{flalign}
\end{subequations}
with $\Delta_{\rm p}=\omega_{\rm p}-\left(\omega_{2}-\omega_1\right)~,$ $\Delta_{\rm c}=\omega_{\rm c}-\left(\omega_{1}-\omega_0\right)~,$ and $\Delta_{\rm R}=\Delta_{\rm p}+\Delta_{\rm c}~$. Here $\Delta_{\rm p,(c)}$ is the detuning of probe (coupling) laser, $\omega_{\rm p, (c)}$ is the angular frequency of probe (coupling) laser, $\Delta_{\rm R}$ is the two-photon Raman detuning, and h.c. is the hermitian conjugate. The first term of the total Hamiltonian, ${\cal H}_0$, represents the field-free atomic system, whereas the second term of the Hamiltonian, ${\cal H}_{\rm I}$, describes the interaction with both probe and coupling fields.

Using standard semiclassical methods \cite{berman}, the equations of motion for the density matrix elements, $\rho_{ij}$, are \begin{subequations}\label{eq:blocheq}
\begin{eqnarray}
\dot{\rho}_{00}=&&\Gamma_{1}\rho_{11}+\frac{{\rm i}\Omega_{\text{c}}}{2}(\rho_{01}-\rho_{10}),\label{eq:blocheq1}\\
\dot{\rho}_{11}=&-&\Gamma_{1}\rho_{11}+\Gamma_{2}\rho_{22}-\frac{{\rm i}\Omega_{\text{c}}}{2}(\rho_{01}-\rho_{10})\nonumber\\
&+&\frac{{\rm i}\Omega_{\text{p}}}{2}(\rho_{12}-\rho_{21}),\label{eq:blocheq2}\\
\dot{\rho}_{22}=&-&\Gamma_{2}\rho_{22}-\frac{{\rm i}\Omega_{\text{p}}}{2}(\rho_{12}-\rho_{21}),\label{eq:blocheq3}\\
\dot{\rho}_{01}=&-&({\rm i}\Delta_{\text{c}}+\gamma')\rho_{01}-\frac{{\rm i}\Omega_{\text{c}}}{2}(\rho_{11}-\rho_{00})\nonumber\\
&+&\frac{{\rm i}\Omega_{\text{p}}}{2}\rho_{02},\label{eq:blocheq4}\\
\dot{\rho}_{12}=&-&({\rm i}\Delta_{\text{p}}+\gamma'')\rho_{12}-\frac{{\rm i}\Omega_{\text{p}}}{2}(\rho_{22}-\rho_{11})\nonumber\\
&-&\frac{{\rm i}\Omega_{\text{c}}}{2}\rho_{02},\label{eq:blocheq5}\\
\dot{\rho}_{02}=&-&({\rm i}\Delta_{\text{R}}+\gamma''')\rho_{02}+\frac{{\rm i}\Omega_{\text{p}}}{2}\rho_{01}\nonumber\\
&-&\frac{{\rm i}\Omega_{\text{c}}}{2}\rho_{12},\label{eq:blocheq6}
 \end{eqnarray}
\end{subequations}
where we define effective linewidths $\gamma'=\Gamma_{1}/2+\gamma_{\rm c} $, $\gamma'' = (\Gamma_{1}+\Gamma_{2})/2+\gamma_{\rm p}~$, and $\gamma'''=\Gamma_{2}/2+\gamma_{\rm p}+\gamma_{\rm c}~$ and $\Gamma_1$ and $\Gamma_2$ are the natural linewidths of the states $\ket{1}$ and $\ket{2}$ respectively. In addition to spontaneous decay, we include a dephasing due to the linewidth of the probe and coupling fields of $\gamma_{\rm p}$ and $\gamma_{\rm c}$, respectively.
Solving Eqs. (\ref{eq:blocheq}) (with $\dot{\rho}_{ij}=0$) together with the constraint $\rho_{00}+\rho_{11}+\rho_{22}=1$ using a perturbation technique (see Appendix \ref{sec:app1}), the steady state solutions of the density matrix $\rho_{ij}$ are given by,

\begin{widetext}
\begin{subequations}
\begin{eqnarray}
\rho_{01}&=&\frac{{\rm i}\Omega_{\rm c}}{2}\left[\gamma'+{\rm i}\Delta_{\rm c}+\frac{\Omega_{\rm c}^2\gamma'/\Gamma_1}{\gamma'-{\rm i}\Delta_{\rm c}}\right]^{-1}~,\label{eq:steady1}\\
\rho_{11}&=&\frac{\Omega_{\rm c}^2\gamma'/2}{\Gamma_1\Delta_{\rm c}^2+\Gamma_1\gamma'^2+\gamma'\Omega_{\rm c}^2}~,\label{eq:steady2}\\
\rho_{02}&=&\frac{2\Gamma_1({\rm i}\Delta_{\rm p}+\gamma'')({\rm i}\Delta_{\rm c}-\gamma')\Omega_{\rm c}\Omega_{\rm p}+\gamma'\Omega_{\rm c}^3\Omega_{\rm p}}{2(\Gamma_1\Delta_{\rm c}^2+\Gamma_1\gamma'^2+\gamma'\Omega_{\rm c}^2)[4({\rm i}\Delta_{\rm p}+\gamma'')({\rm i}\Delta_{\rm R}+\gamma''')+\Omega_{\rm c}^2]}~,\label{eq:steady3}\\
\rho_{12}&=&\frac{{\rm i}\Omega_{\rm c}^2\Omega_{\rm p}\gamma'/4}{\Gamma_1\Delta_{\rm c}^2+\Gamma_1\gamma'^2+\gamma'\Omega_{\rm c}^2}\left[1+\frac{\gamma_{\rm c}(1+{\rm i}\Delta_{\rm c}/\gamma')}{\gamma''+{\rm i}\Delta_{\rm p}}\right]\left[\gamma'''+{\rm i}\Delta_{\rm R}+\frac{\Omega_{\rm c}^2/4}{\gamma''+{\rm i}\Delta_{\rm p}}\right]^{-1}~.\label{eq:steady4}
\end{eqnarray}
\end{subequations}
\end{widetext}
In the weak excitation limit of $\Omega_{\rm p} \ll \gamma''$, $\rho_{22}=0$. We can therefore assume that no population is lost from the system via other decay channels from state $|2\rangle$, although additional decay channels may contribute to the linewidth, $\Gamma_2$.

We consider $\rho_{12}$ as it determines the complex susceptibility of the system. It is clear from Eq. (3d) that the number of atoms pumped into state $|1\rangle$ strongly affects the magnitude of $\rho_{12}$, manifest as the multiplication factor $\rho_{11}\Omega_{\rm p}/2$. In our case, the population of state $|1\rangle$ is resonantly pumped from state $\ket{0}$, i.e., $\Delta_{\rm c}\approx0$, and $\gamma_{\rm c}$ is much less than $\gamma''$. Thus, the second term in the first square bracket approaches unity within this approximation.


\subsection{The complex susceptibility of the system}
The complex susceptibility of the system at the probe frequency is obtained by comparing the polarization obtained from classical electrodynamics with that calculated using a density matrix treatment \cite{bana95,eit_review}. The expression of the complex susceptibility is then given by \cite{monsit}
\begin{equation}
\chi=-\frac{2{\cal N}d_{21}^2}{\hbar\epsilon_0\Omega_{\rm p}}\rho_{21}~,
\end{equation}
where ${\cal N}$ is the atomic density and $d_{21}$ is the dipole matrix element for probe transition. For $|\chi|<1$ the real part and imaginary part of the susceptibility are respectively proportional to the refractive index $n_{\rm R}$ and the absorption coefficient $\alpha$ via the relations
\begin{subequations}
\begin{eqnarray}
n_{\rm R}&=&1+{\rm Re}[\chi]/2~,\label{eq:rel1}\\
\alpha &=& k_{\rm p}{\rm Im}[\chi]~,
\end{eqnarray}
\end{subequations}
where $k_{\rm p}$ is the wavevector of the probe field.
Thus the complex susceptibility of the system is
\begin{equation}
\chi=\frac{{\rm i}{\cal N}_1 d_{21}^2}{\hbar \epsilon_0}\left[\gamma'''-{\rm i}\Delta_{\rm R}+\frac{\Omega_{\rm c}^2/4}{\gamma''-{\rm i}\Delta_{\rm p}}\right]^{-1}~.\label{eq:comchi}
\end{equation}
Here ${\cal N}_1 = {\cal N} \rho_{11}$, the atomic density of state $|1\rangle$. According to Eq. (6), the complex susceptibility is proportional to the number of the atoms in state $|1\rangle$. As state $|1\rangle$ is coherently pumped from state $\ket{0}$, the multiplication factor has a Lorentzian profile as a function of $\Delta_{\rm c}$. This implies that to get a large susceptibility, one needs to resonantly pump population to state $|1\rangle$. The term in the square bracket is similar to the result obtained for EIT in the three-level cascade system~\cite{bana95}. It indicates that, for a finite value of $\Omega_{\rm c}$, the system will become transparent when the probe field is scanned across the resonance at $\Delta_{\rm p}=0$ due to Autler-Townes splitting \cite{autl55}.
When state $|1\rangle$ is weakly pumped by the coupling field, i.e., $\Omega_{\rm c}^2\ll\Gamma_2(\Gamma_2+\Gamma_1)$, the term in the square bracket of Eq. (\ref{eq:comchi}) can be expanded using a Taylor expansion. Neglecting the higher order terms in $\Omega_{\rm c}$, the complex susceptibility reduces to
\begin{equation}
\chi=\frac{{\rm i}{\cal N}_1 d_{21}^2}{\hbar \epsilon_0}\left[\frac{1}{\gamma'''-{\rm i}\Delta_{\rm R}}\right]~.\label{eq:twolike}
\end{equation}
This complex susceptibility is similar to that of a two-level system, except for the multiplication factor. The transmission lineshape of the system is simply a Lorentzian centered at $-\Delta_{\rm c}$ with FWHM $\gamma'''$. For a sufficiently small laser linewidth compared to $\Gamma_2$, the approximation of the FWHM is solely determined by the linewidth of the excited state, $\Gamma_2$, irrespective of the linewidth of state $|1\rangle$, $\Gamma_1$, i.e.,
\begin{equation}
\Gamma_{\rm FWHM}=\Gamma_2~.
\end{equation}
This result is different to the case in which state $|1\rangle$ is incoherently populated. In such a case, the FWHM of the transmission lineshape is determined by the sum of the linewidths from both the lower state and the upper state, i.e., $\Gamma_1+\Gamma_2$ \cite{bransden}.

For experiments in room temperature vapors, the Doppler effect must be included into the model and this topic will be discussed in the next section.


\section{Effect of Doppler broadening}
Each atomic velocity class in a thermal vapor experiences a different laser detuning $\Delta_{\rm p}$ and $\Delta_{\rm c}$ due to the Doppler effect. To obtain the velocity-dependent complex susceptibility, we make changes to $\Delta_{\rm p}$, $\Delta_{\rm c}$ and ${\cal N}$ with the following substitutions \cite{bana95}:
\begin{subequations} \label{eq:dopplersubs}
\begin{eqnarray}
\Delta_{\rm p}&\to&\Delta_{\rm p}-k_{\rm p}v~,\\
\Delta_{\rm c}&\to&\Delta_{\rm c}+k_{\rm c}v~,\\
{\cal N}&\to&\frac{{\cal N}}{u\sqrt{\pi}}{\rm exp}\left(-\frac{v^2}{u^2}\right)~,
\end{eqnarray}
\end{subequations}
where $k_{\rm p(c)}$ is the wavevector of the probe (coupling) field, $u=\sqrt{2k_{\rm B}T/m}$ is the most probable speed of the atoms at a given temperature $T$ and $m$ is the mass of an atom. Substituting Eqs. (\ref{eq:dopplersubs}) into Eq. (\ref{eq:comchi}), the value of the complex susceptibility of a particular velocity class $v$ is then given by,
\begin{eqnarray}
\chi(v){\rm d}v&=&-\frac{{\cal N}d_{21}^2\Omega_{\rm c}^2}{\hbar\epsilon_0 \sqrt{\pi}k_{\rm c}^2(k_{\rm c}-k_{\rm p})u^3}\frac{\gamma^{\prime}}{2\Gamma_1}\left\{\frac{{\rm e}^{-z^2}}{(z+\beta)^2+\sigma^2}\right\}\nonumber\\
&\times&\left[z-z_0+\frac{\Omega_{\rm c}^2/4}{(k_{\rm c}-k_{\rm p})k_{\rm p}u^2(z-z_1)}\right]^{-1}{\rm d}z~,\label{eq:cv}
\end{eqnarray}
with the change of variable $z=v/u$, and
\begin{subequations}
\begin{eqnarray}
\gamma&=&\frac{\gamma'''}{(k_{\rm c}-k_{\rm p})u}~,\\
\sigma&=&\frac{1}{k_{\rm c}u}\sqrt{\gamma^{'2}+\frac{\Omega_{\rm c}^2\gamma'}{\Gamma_1}}~,\\
\xi&=&\frac{\Delta_{\rm R}}{(k_{\rm c}-k_{\rm p})u}~,\\
\beta&=&\frac{\Delta_{\rm c}}{k_{\rm c}u}~,\\
z_0&=&-\xi-{\rm i}\gamma~,\\
z_1&=&\frac{\Delta_{\rm p}+{\rm i}\gamma''}{k_{\rm p}u}~.
\end{eqnarray}
\end{subequations}
The total susceptibility is obtained by integrating Eq. (\ref{eq:cv}) over all velocity classes. The full result of the integration is discussed in Appendix B. We consider the case in which the coupling Rabi frequency, $\Omega_{\rm c}$ is sufficiently weak that the EIT-like third term in the square bracket of Eq. (\ref{eq:cv}) is neglected. In this case the complex susceptibility, $\chi_{\rm D}$, becomes
\begin{flalign}\label{eq:cd}
\chi_{\rm D}(\Delta_{\rm p})&=-\frac{{\cal N}d_{21}^2\Omega_{\rm c}^2}{\hbar\epsilon_0\sqrt{\pi}k_{\rm c}^2(k_{\rm c}-k_{\rm p})u^3}\frac{\gamma'}{2\Gamma_1}\times\nonumber\\
&\int_{-\infty}^{\infty}\left\{\frac{{\rm e}^{-z^2}}{(z+\beta)^2+\sigma^2}\right\}\left[\frac{1}{(z+\xi)+{\rm i}\gamma}\right]{\rm d}z~.
\end{flalign}
From Eq. (\ref{eq:cd}), the total complex susceptibility is simply the convolution between a Lorentzian of width $\gamma$ (term in square bracket describing the transition from lower state $|1\rangle$ to upper state $|2\rangle$) and a product of a Lorentzian of width $\sigma$ and a Gaussian (term in curly bracket describing the atomic velocity distribution of lower state $|1\rangle$).

The result of the integration in Eq. (\ref{eq:cd}) involves the Faddeva function \cite{siddon2} (the exact result of the integration is described in Appendix C). However, the integration can be simplified by replacing the product between a Gaussian and a Lorentzian with a Lorentzian, i.e.,
\begin{equation}
\frac{{\rm e}^{-z^2}}{(z+\beta)^2+\sigma^2}\to\frac{{\rm e}^{-\beta^2}}{(z+\beta)^2+\sigma^2}.
\end{equation}
This approximation is valid since, at room temperature, the width of the Lorentzian $\sigma$ is much smaller than the width of the Gaussian. Hence the product of the Gaussian and the Lorentzian approximately vanishes when $|z|>\sigma$ and we can approximate the product by the Lorentzian of width $\sigma$. In other words, the range of velocity classes involved in the integration around the position where the Lorentzian of width $\sigma$ is centered is much smaller than the most probable speed of the atoms, $v\ll u$. Hence, we can expand the Gaussian around the position where the Lorentzian is centered, i.e., ${\rm exp(-z^2)}\approx{\rm exp(-\beta^2)}$.

Using this approximation, the total susceptibility is simply given by
\begin{equation}
\chi_{\rm D}(\Delta_{\rm p})=-\frac{{\cal N}d_{21}^2\Omega_{\rm c}^2\sqrt{\pi}}{\hbar\epsilon_0 k_{\rm c}^2u\sigma}\frac{\gamma'}{2\Gamma_1}\frac{1}{\Delta_{\rm R}+{\rm i}\Gamma_{\rm FWHM}/2}~,
\end{equation}
with
\begin{equation}
\frac{\Gamma_{\rm FWHM}}{2}=\gamma'''+\left(\frac{k_{\rm c}-k_{\rm p}}{k_{\rm c}}\right)\sqrt{\gamma'^2+\frac{\Omega_{\rm c}^2\gamma'}{\Gamma_1}}~.
\end{equation}

It is clear from Eqs. (14) and (15) that the absorption lineshape remains Lorentzian with the FWHM of $\Gamma_{\rm FWHM}$ even when the Doppler effect is included. It is worth noting that the total susceptibility in this case is different from the total susceptibility calculated for the case of incoherent pumping. The total susceptibility for incoherent pumping is the convolution between a Lorentzian and a Gaussian, resulting in a Voigt profile \cite{siddon1}.

In the limit where $\Omega_{\rm c}/\Gamma_1\ll1$, the linewidth of the absorption profile is simply,
\begin{equation}
\Gamma_{\rm FWHM}=\Gamma_2+\left(\frac{k_{\rm c}-k_{\rm p}}{k_{\rm c}}\right)\Gamma_1,
\end{equation}
(neglecting $\gamma_{\rm p}$ and $\gamma_{\rm c}$). It contains the sum of two terms: the first term is the linewidth of the absorption lineshape in the case in which the Doppler effect is neglected and the latter is the linewidth of the lower state scaled by the ratio obtained from the wavevectors. Physically, the second term originates from the fact that the atoms are velocity-selected by the Doppler effect for the atom-field interaction. Only atoms whose velocities are between $-\Gamma_1/2k_{\rm c}$ and $\Gamma_1/2k_{\rm c}$ are coherently pumped into the lower state when the coupling field is on resonance. Since $\Gamma_1/k_{\rm c}$ is very small compared to the width of the Doppler broadening, all atoms pumped into state $|1\rangle$ have approximately the same velocity distribution, i.e., the distribution is independent of velocity, and given by
\begin{equation}
f(v)=\frac{1}{u\sqrt{\pi}}{\rm e}^{-\beta^2}~.
\end{equation}
However, the distribution of the atoms is also determined by the Lorentzian of width $\Gamma_1/k_{\rm c}$. Thus the final velocity distribution of the pumped atoms is a Lorentzian of width $\Gamma_1/k_{\rm c}$ and the height is scaled by Eq. (17). In the two-photon interaction process, the width of $\Gamma_1/k_{\rm c}$ in velocity space is equivalent to the width of $(k_{\rm c}-k_{\rm p})\Gamma_1/k_{\rm c}$ in frequency space. Hence the total linewidth of the final absorption lineshape is the sum of $(k_{\rm c}-k_{\rm p})\Gamma_1/k_{\rm c}$ with the unaffected linewidth, $\Gamma_2$.


\section{Comparison between theory and experimental results}
\begin{figure}[h!]
\begin{center}
\includegraphics[width=8.4cm]{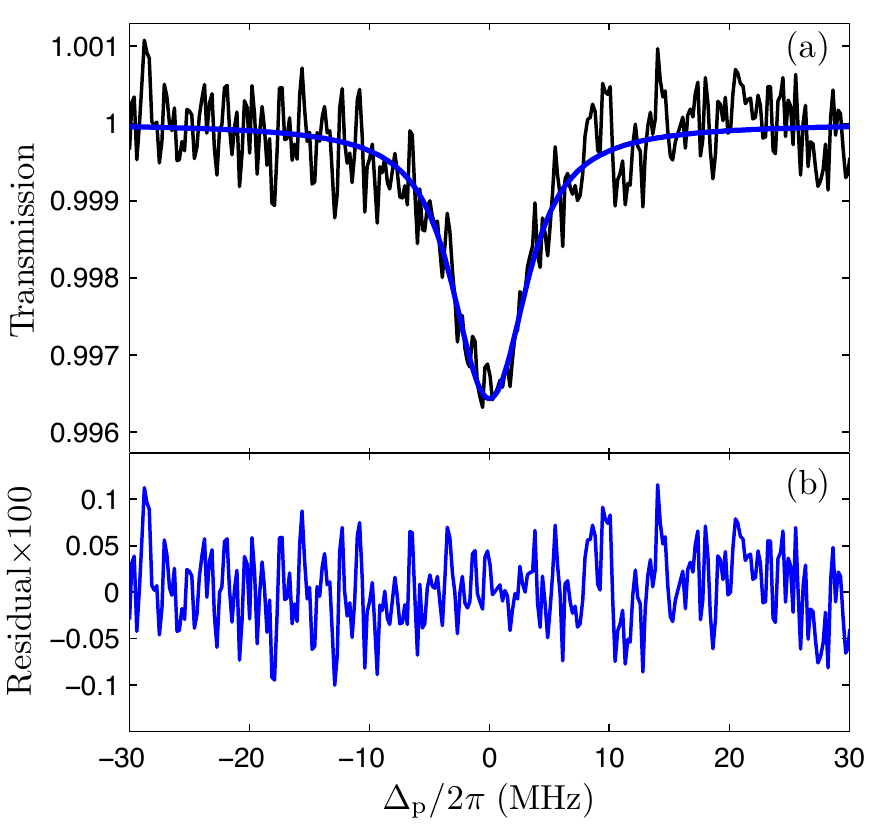}
\caption[]{(Color online) (a) Comparison between the observed spectra, shown by a black solid line, for a relatively small $\Omega_{\rm c}^{\rm r}$ = 0.6 MHz, and the theoretical model.
The grey (blue) solid line is the theoretical prediction, taking into account the absorption of the coupling field across the vapor cell. (b) The residual plot between the observed data and the theoretical model.\label{fig2}}
\end{center}
\end{figure}

\begin{figure}[t!]
\begin{center}
\includegraphics[width=8.4cm]{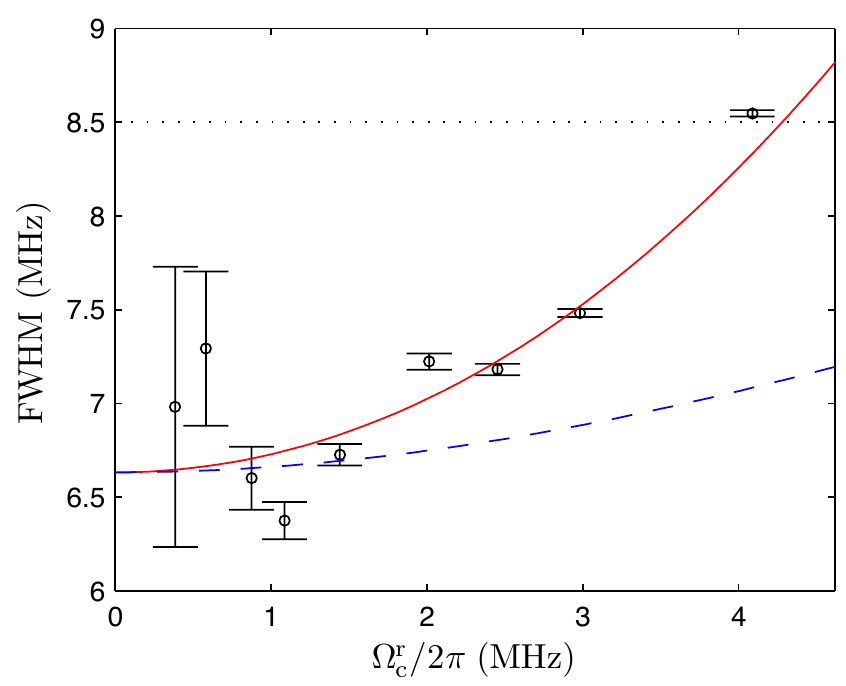}
\caption[]{(Color online) The values of the FWHM are plotted as a function of $\Omega_{\rm c}^{\rm r}$. The dashed curve is the theoretical prediction calculated using the weak coupling model, whereas the solid (red) curve is the theoretical prediction calculated using the full model. As expected the data are in agreement with the weak coupling model when $\Omega_{\rm c}^{\rm r}$ is relatively small. However, for a large $\Omega_{\rm c}^{\rm r}$, the weak coupling model fails to predict the FWHM. This difference at large $\Omega_{\rm c}^{\rm r}$ can be recovered when using the full model. The combined lower and upper state linewidth $\Gamma_1 + \Gamma_2$ is shown by the dotted line. \label{fig3}}
\end{center}
\end{figure}
 To test the result derived above we used the experimental setup described by Carr et al. \cite{carr}. The experiment was performed in a 7.5~cm vapor cell containing Cs at room temperature. The 1470 nm weak probe beam (with horizontal linear polarization) and the 852 nm co-axial, counter-propagating coupling beam (with horizontal linear polarization) are applied along the vapor cell axis. The probe and coupling beams have $1/e^2$ radii of 1.2 mm and 1.6 mm, respectively. The coupling beam was stabilised to the 6S$_{1/2},F=4\to~$6P$_{3/2},F'=5$ transition while the probe beam was scanned across the 6P$_{3/2},F'=5\to~$7S$_{1/2},F''=4$ transition. The scan is calibrated using a wavemeter to better than 1\% accuracy. The transmission signal measured from the experiment is shown as the solid black line in Fig. \ref{fig2}(a).

To model the transmission lineshape, the complex susceptibility is calculated for each magnetic sublevel and the total complex susceptibility is the average of all complex susceptibilities over all possible magnetic sublevels. This is given by
\begin{equation}
\chi_{\rm TOT}(\Delta_{\rm p})=\frac{1}{16}\sum_{m_{F}=-4}^{4}\chi_{\rm D}^{m_F}(\Delta_{\rm p})~,
\end{equation}
where $\chi_{\rm D}^{m_F}(\Delta_{\rm p})$ is the complex susceptibility corresponding to the $m_F$ magnetic sublevel of 6S$_{1/2},F=4$ state. $\chi_{\rm D}^{m_F}$ is calculated using Eq. (14), where the coupling Rabi frequency is sufficiently weak. The factor of 1/16 in the equation accounts for the fact that the initial population is evenly distributed among the magnetic sublevels of 6S$_{1/2},F=(3,4)$. The coupling Rabi frequency of the transition 6S$_{1/2},F=4\to~$6P$_{3/2},F'=5$ and the dipole matrix element of the transition 6P$_{3/2},F'=5\to~$7S$_{1/2},F''=4$ corresponding to each magnetic sublevel, $\Omega_{\rm c}^{m_F}$ and $d_{21}^{m_F}$, are given by
\begin{subequations}
\begin{flalign}\label{eq:reducedRF}
\Omega_{\rm c}^{m_F}&=\Omega_{\rm c}^{\rm r}\times\sqrt{11}\left( \begin{array}{ccc}5 & 1 & 4\\m_{F'} & 0 & -m_{F}\end{array}\right)~,\\
d_{21}^{m_F}&=5.63ea_0\times\sqrt{11/3}\left( \begin{array}{ccc}5 & 1 & 4\\m_{F'} & 0 & -m_{F}\end{array}\right)~,
\end{flalign}
\end{subequations}
where the reduced dipole matrix element of the transition 6S$_{1/2},F=4\to$6P$_{3/2},F'=5$ is absorbed into $\Omega_{\rm c}^{\rm r}$, i.e., $\Omega_{\rm c}^{\rm r}\equiv eE_{\rm c}\langle$6P$_{3/2}\|r\|$6S$_{1/2}\rangle/\hbar$, $a_0$ is Bohr radius, and, $m_F$ and $m_{F'}$ are the magnetic sublevels of the 6S$_{1/2}$ and 6P$_{3/2}$, respectively.

The comparison between the experimental data and the theoretical model is shown in Fig. 2(a). The theoretical transmission shown as the solid grey (blue) line was calculated using Eqs. (5b) and (18) for a known temperature of $T=22$ $^\circ$C. The fit parameters $\Omega_{\rm c}^{\rm r}/2\pi=0.6$ MHz, $\gamma_{\rm p}/2\pi=0.2$ MHz are in good agreement with the experimental values taking into account the variation of the coupling intensity across the vapor cell. The residual plot in Fig. 2(b) shows that the theoretical model is in good agreement with the experimental data. Fig. 3 shows the transmission lineshape FWHM as a function $\Omega_{\rm c}^{\rm r}$. The dashed line shows the FWHM as a function of $\Omega_{\rm c}^{\rm r}$ for the weak coupling field approximation. The dependence on $\Omega_{\rm c}^{\rm r}$ shows good agreement when $\Omega_{\rm c}^{\rm r}$ is small. However when $\Omega_{\rm c}^{\rm r}$ is large, the disagreement between the model and the experimental data increases. The agreement between the theory and the experiment can be recovered when the complete solution (see Appendix B) is used to calculate the FWHM as a function of $\Omega_{\rm c}^{\rm r}$ (shown as the solid line). Note that both curves approach the same value of $2\pi~\times$~6.6 MHz when $\Omega_{\rm c}^{\rm r}$ approaches zero. This is sub-natural and less than the combined linewidth $\Gamma_1+\Gamma_2 5.2+3.3=8.5$~MHz \cite{theo84} (shown as the dot dash horizontal line). The difference between the model and the experiment at large $\Omega_{\rm c}^{\rm r}$ arises from Autler-Townes splitting \cite{autl55} and is described by the EIT-like term in Eq. (6). The comparison  between experiment and theory for large $\Omega_{\rm c}^{\rm r}$ will be the topic of the next section.


\section{Comparison between experiment and theory for large $\Omega_{\rm c}^{\rm r}$}
\begin{figure}[t!]
\begin{center}
\includegraphics[width=8.4cm]{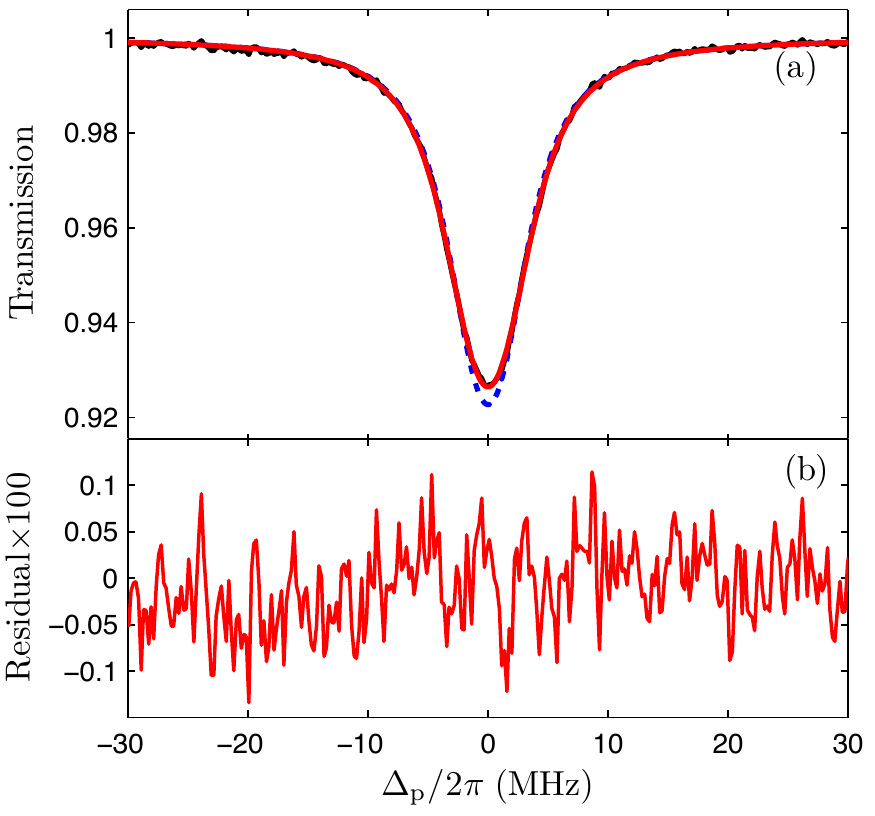}
\caption[]{(Color online) (a) Comparison between the observed spectra, shown by a black solid line, for $\Omega_{\rm c}^{\rm r}/2\pi=3.0$ MHz, and the theoretical model.
The grey (blue) dotted line is the theoretical prediction calculated using the weak coupling model, whereas the grey (red) solid line is the theoretical prediction calculated using the full model. (b) The residual plot between the observed data and the theoretical model calculated using the full model.\label{fig4}}
\end{center}
\end{figure}
We now apply the same method to model the transmission lineshape, except that we are no longer in the weak pumping regime and the third term in Eq. (10) can no longer be neglected. Eq. (B8) is now used to calculate the complex susceptibility.

Fig.~4(a) shows the comparison between the observed transmission lineshape, shown by a black solid line, and the theoretical transmission lineshapes calculated using both the weak coupling approximation, shown by a grey (blue) dashed line, and the full model, shown by a grey (red) solid line. The theoretical curves were calculated with $\Omega_{\rm c}^{\rm r}/2\pi=3.0$ MHz. The transmission calculated using the full model shows good agreement with observed data, resulting in the small residual as shown in Fig. \ref{fig4}(b). It is observed that the weak coupling approximation is in good agreement with the observed data, except at the region around the resonance.

Fig.~5(a) shows the transmission lineshape, shown by a black solid line, in a case where $\Omega_{\rm c}^{\rm r}$ is large enough to Autler-Townes split the absorption resonance. The value of $\Omega_{\rm c}^{\rm r}/2\pi$ in this case is 15 MHz. The grey (red) solid line is the theoretical prediction calculated using the full model in the region $|\Delta_{\rm p} / 2\pi|\lesssim20$~MHz. Both the theoretical prediction and the observed data are in good agreement around resonance. Fig.~5(b) shows the residual between the theoretical prediction and the observed data.
\begin{figure}[t!]
\begin{center}
\includegraphics[width=8.4cm]{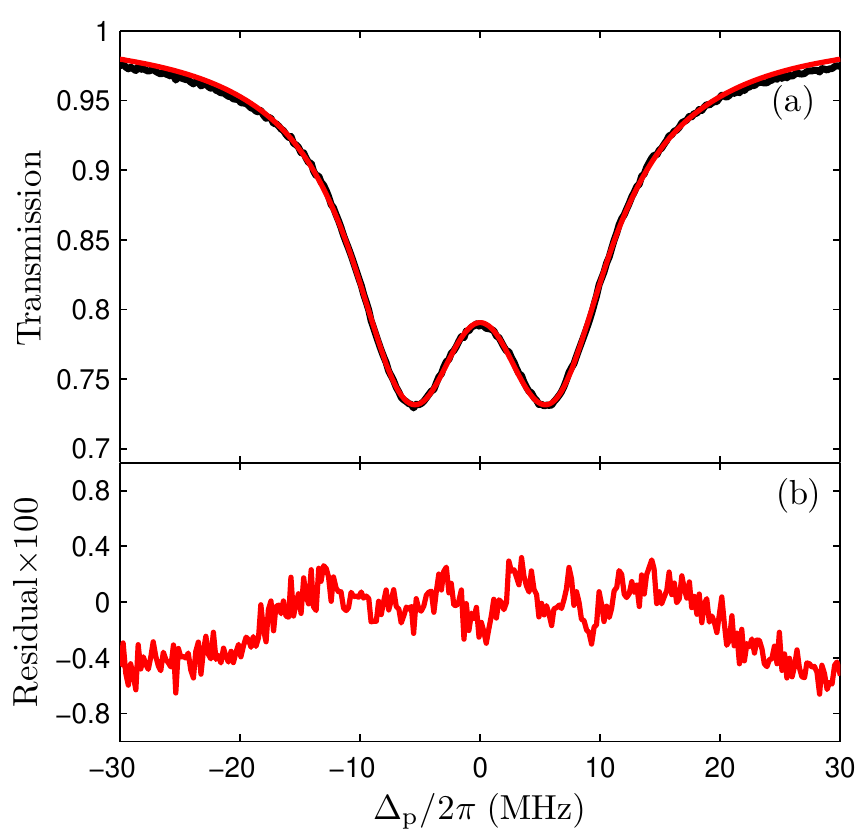}
\caption[]{(Color online) (a) Comparison between the observed spectra, shown by a black solid line, for $\Omega_{\rm c}^{\rm r}/2\pi=15$ MHz, and the theoretical model.
The grey (red) solid curve is the theoretical prediction calculated using the full model. (b) The residual plot between the observed data and the theoretical model.\label{fig5}}
\end{center}
\end{figure}


\section{Conclusion}

We have developed the theory of absorption lineshapes on excited state transitions where the lower state is coherently populated. We show that for an atom at rest, in the limit of weak pumping, the lineshape is a Lorentzian and the linewidth of the transition reduces to the linewidth of the upper state. Including the effect of Doppler broadening the linewidth is still subnatural and we find that the predicted lineshape is in very good agreement with experimental data over a wide range of coupling field parameters.


\section*{Acknowledgments}
We thank R. M. Potvleige and I. G. Hughes for discussions and the UK Engineering and Physical Sciences Research Council for financial support.


\appendix
\section{Steady state solutions by perturbation technique}\label{sec:app1}
Since the probe field is sufficiently weak (it will be shown later that the weak probe condition is fulfilled when $\Omega_{\rm p}/\gamma'\ll1$), one can consider the expansion of the density matrix, $\rho_{ij}$, in the power of $\Omega_{\rm p}$, namely,
\begin{equation}\label{eq:pert}
\rho_{ij}=\rho_{ij}^{(0)}+\rho_{ij}^{(1)}\Omega_{\rm p}+\rho_{ij}^{(2)}\Omega_{\rm p}^{2}+\rho_{ij}^{(3)}\Omega_{\rm p}^{3}+\ldots~,
\end{equation}
where $\rho_{ij}^{(n)}$ is the $n^{\rm th}$ order correction of the expansion of $\rho_{ij}$. To solve for $\rho_{ij}$, we substitute Eq. (\ref{eq:pert}) into Eqs. (\ref{eq:blocheq}), equate the terms of the same power in $\Omega_{\rm p}$, and then solve for $\rho_{ij}^{(n)}$ from $n=0$ to all $n$. \cite{nayfeh}.

Applying this technique to Eqs. (\ref{eq:blocheq}), the set of equations corresponding to the zeroth power of $\Omega_{\rm p}$ is given by
\begin{subequations}\label{eq:pert0}
\begin{flalign}
&\Gamma_1\rho_{11}^{(0)}+\frac{\rm i}{2}\Omega_{\rm c}\left(\rho_{01}^{(0)}-\rho_{10}^{(0)}\right)=0~,\label{eq:pert01}\\
&\Gamma_{2}\rho_{22}^{(0)}=0~,\label{eq:pert02}\\
&\left({\rm i}\Delta_{\rm c}+\gamma'\right)\rho_{01}^{(0)}+\frac{\rm i}{2}\Omega_{\rm c}\left(\rho_{11}^{(0)}-\rho_{00}^{(0)}\right)=0~,\label{eq:pert03}\\
&\left({\rm i}\Delta_{\rm p}+\gamma''\right)\rho_{12}^{(0)}+\frac{\rm i}{2}\Omega_{\rm c}\rho_{02}^{(0)}=0~,\label{eq:pert04}\\
&\left({\rm i}\Delta_{\rm R}+\gamma'''\right)\rho_{02}^{(0)}+\frac{\rm i}{2}\Omega_{\rm c}\rho_{12}^{(0)}=0~,\label{eq:pert05}\\
&\rho_{00}^{(0)}+\rho_{11}^{(0)}+\rho_{22}^{(0)}=1~.\label{eq:pert06}
\end{flalign}
\end{subequations}
 We find that the zeroth order corrections of $\rho_{ij}$ vanish, except $\rho_{00}^{(0)}$, $\rho_{11}^{(0)}$ and $\rho_{01}^{(0)}$. The expression for $\rho_{01}^{(0)}$ and $\rho_{11}^{(0)}$ are given by
\begin{subequations}\label{zeroth}
\begin{eqnarray}
\rho_{01}^{(0)}&=&\frac{{\rm i}\Omega_{\rm c}}{2}\left[\gamma'+{\rm i}\Delta_{\rm c}+\frac{\Omega_{\rm c}^2\gamma'/\Gamma_1}{\gamma'-{\rm i}\Delta_{\rm c}}\right]^{-1}~,\label{eq:zeroth1}\\
\rho_{11}^{(0)}&=&\frac{\Omega_{\rm c}^2\gamma'/2}{\Gamma_1\Delta_{\rm c}^2+\Gamma_1\gamma'^2+\gamma'\Omega_{\rm c}^2}~.\label{eq:zeroth2}
\end{eqnarray}
\end{subequations}
Similarly, the set of equations corresponding to the $n^{\rm th}$ power of $\Omega_{\rm p}$ (for $n\ge1$) is given by
\begin{subequations}\label{eq:pertn}
\begin{flalign}
&\Gamma_1\rho_{11}^{(n)}+\frac{\rm i}{2}\Omega_{\rm c}\left(\rho_{01}^{(n)}-\rho_{10}^{(n)}\right)=0~,\label{eq:pertn1}\\
&\Gamma_{2}\rho_{22}^{(n)}+\frac{\rm i}{2}\left(\rho_{12}^{(n-1)}-\rho_{21}^{(n-1)}\right)=0~,\label{eq:pertn2}\\
&\left({\rm i}\Delta_{\rm c}+\gamma'\right)\rho_{01}^{(n)}-\frac{\rm i}{2}\rho_{02}^{(n-1)}\nonumber\\
&~~~~~~~~~~~~~~~~~~~~~~+\frac{\rm i}{2}\Omega_{\rm c}\left(\rho_{11}^{(n)}-\rho_{00}^{(n)}\right)=0~,\label{eq:pertn3}\\
&\left({\rm i}\Delta_{\rm p}+\gamma''\right)\rho_{12}^{(n)}+\frac{\rm i}{2}\Omega_{\rm c}\rho_{02}^{(n)}\nonumber\\
&~~~~~~~~~~~~~~~~~~~~~+\frac{\rm i}{2}\left(\rho_{22}^{(n-1)}-\rho_{11}^{(n-1)}\right)=0~,\label{eq:pertn4}\\
&\left({\rm i}\Delta_{\rm R}+\gamma'''\right)\rho_{02}^{(n)}-\frac{\rm i}{2}\rho_{01}^{(n-1)}+\frac{\rm i}{2}\Omega_{\rm c}\rho_{12}^{(n)}=0~,\label{eq:pertn5}\\
&\rho_{00}^{(n)}+\rho_{11}^{(n)}+\rho_{22}^{(n)}=0~.\label{eq:pertn6}
\end{flalign}
\end{subequations}
Using Eqs. (\ref{zeroth}) and Eqs. (\ref{eq:pertn}), all of $\rho_{ij}^{(1)}$ again vanish, except $\rho_{02}^{(1)}$ and $\rho_{12}^{(1)}$, whose expressions are given by,
\begin{widetext}
\begin{subequations}
\begin{eqnarray}
\rho_{02}^{(1)}&=&\frac{2\Gamma_1({\rm i}\Delta_{\rm p}+\gamma')({\rm i}\Delta_{\rm c}-\gamma')\Omega_{\rm c}+\gamma'\Omega_{\rm c}^3}{2(\Gamma_1\Delta_{\rm c}^2+\Gamma_1\gamma'^2+\gamma'\Omega_{\rm c}^2)[4({\rm i}\Delta_{\rm p}+\gamma'')({\rm i}\Delta_{\rm R}+\gamma''')+\Omega_{\rm c}^2]}~,\label{eq:first1}\\
\rho_{12}^{(1)}&=&\frac{{\rm i}\Omega_{\rm c}^2\gamma'/4}{\Gamma_1\Delta_{\rm c}^2+\Gamma_1\gamma'^2+\gamma'\Omega_{\rm c}^2}\left[1+\frac{\gamma_{\rm c}(1+{\rm i}\Delta_{\rm c}/\gamma')}{\gamma''+{\rm i}\Delta_{\rm p}}\right]\left[\gamma'''+{\rm i}\Delta_{\rm R}+\frac{\Omega_{\rm c}^2/4}{\gamma''+{\rm i}\Delta_{\rm p}}\right]^{-1}~.\label{eq:first2}
\end{eqnarray}
\end{subequations}
\end{widetext}
It can be shown that $\rho_{12}^{(2)}=\rho_{02}^{(2)}=0$. The solutions of the coherence $\rho_{02}$ and $\rho_{12}$, to second order, are then
\begin{subequations}
\begin{eqnarray}
\rho_{02}&=&\Omega_{\rm p}\rho_{02}^{(1)}~,\\
\rho_{12}&=&\Omega_{\rm p}\rho_{12}^{(1)}~.
\end{eqnarray}
\end{subequations}

Thus far, we have assumed only that the probe Rabi frequency is sufficiently weak without quantifying this condition. To quantitatively determine the weak probe condition, let us consider the steady state of $\rho_{01}$ as this quantity is strongly related to $\rho_{11}$ and hence also to $\rho_{12}$. Using Eq. (\ref{eq:blocheq4}) with the substitution $\rho_{00}=1-\rho_{11}-\rho_{22}$, the expression of the steady state of $\rho_{01}$ is given by,
\begin{flalign}
&\rho_{01}=-\frac{{\rm i}\Omega_{\rm c}\rho_{11}}{{\rm i}\Delta_{\rm c}+\gamma'}-\frac{{\rm i}\Omega_{\rm c}\rho_{22}/2}{{\rm i}\Delta_{\rm c}+\gamma'}\nonumber\\
&~~~~~~~~~~~~~~~+\frac{{\rm i}\Omega_{\rm c}/2}{{\rm i}\Delta_{\rm c}+\gamma'}+\frac{{\rm i}\Omega_{\rm p}\rho_{02}/2}{{\rm i}\Delta_{\rm c}+\gamma'}~.
\end{flalign}
The first approximation is to neglect the contribution to $\rho_{01}$ from $\rho_{22}$ as $\rho_{22}$ vanishes up to the second order correction. This approximation is justified as the probe laser is weak and the upper state population is negligible. Thereafter we consider the last term which contains the product between $(\Omega_{\rm p}/2)/({\rm i}\Delta_{\rm c}+\gamma'/2)$ and $\rho_{02}$. The first term is of order $\Omega_{\rm p}/\gamma'$ at resonance $\Delta_{\rm c}=0$. The second term $\rho_{02}$ is also of order $\Omega_{\rm p}/\gamma'$ at resonance (see Eq. (\ref{eq:first1}) and Eq. (7a)). Thus the product of these two terms is of order $(\Omega_{\rm p}/\Gamma_1)^2$, which can be neglected if $\Omega_{\rm p}/\gamma'$ is much less than one. It follows that these approximations lead to the same results for $\rho_{ij}$ given by Eqs. (8). Hence, the weak probe condition is valid when $\Omega_{\rm p}/\gamma'\ll1$.


\section{Complete solution of the Complex susceptibility of the system}
The complete solution of the complex susceptibility can be found by evaluating the integral
\begin{flalign}
&\int_{-\infty}^{\infty}\frac{{\rm e}^{-z^2}}{(z+\beta)^2+\sigma^2}\times\nonumber\\
&~~~~~~~~\left[z-z_0+\frac{\Omega_{\rm c}^2/4}{(k_{\rm c}-k_{\rm p})k_{\rm p}u^2(z-z_1)}\right]^{-1}{\rm d}z~.
\end{flalign}
To evaluate the integral, one re-writes the integrand, using partial fractions, as
\begin{flalign}
&\frac{{\rm e}^{-z^2}}{(z+\beta)^2+\sigma^2}\left[z-z_0+\frac{\Omega_{\rm c}^2/4}{(k_{\rm c}-k_{\rm p})k_{\rm p}u^2(z-z_1)}\right]^{-1}\nonumber\\
=&-\frac{(z_1-\phi_+)}{[(\beta+\phi_+)^2+\sigma^2](\phi_+-\phi_-)}\frac{{\rm e}^{-z^2}}{z-\phi_+}\nonumber\\
&+\frac{(z_1-\phi_-)}{[(\beta+\phi_-)^2+\sigma^2](\phi_+-\phi_-)}\frac{{\rm e}^{-z^2}}{z-\phi_-}\nonumber\\
&-\frac{{\rm i}(z_1+\beta+{\rm i}\sigma)}{2\sigma(\beta+\phi_++{\rm i}\sigma)(\beta+\phi_-+{\rm i}\sigma)}\frac{{\rm e}^{-z^2}}{z+\beta+{\rm i}\sigma}\nonumber\\
&+\frac{{\rm i}(z_1+\beta-{\rm i}\sigma)}{2\sigma(\beta+\phi_+-{\rm i}\sigma)(\beta+\phi_--{\rm i}\sigma)}\frac{{\rm e}^{-z^2}}{z+\beta-{\rm i}\sigma}~,
\end{flalign}
where
\begin{flalign}
\phi_{\pm}=&\frac{1}{2}(z_0+z_1)\nonumber\\
&~~~\pm\frac{1}{2}\sqrt{(z_0-z_1)^2-\frac{\Omega_{\rm c}^2}{k_{\rm p}(k_{\rm c}-k_{\rm p})u^2}}~.
\end{flalign}
The integrand has four poles in the complex plane, i.e., at $\phi_+$, $\phi_-$, $-\beta-{\rm i}\sigma$ and $-\beta+{\rm i}\sigma$. Clearly, this complex integral reduces to the integral of the form
\begin{equation}
\int_{-\infty}^{\infty}\frac{{\rm e}^{-z^2}}{z-z_{\rm p}}{\rm d}z~,
\end{equation}
where $z_{\rm p}$ is the pole in the complex plane. The solution of the integration is given by
\begin{equation}
\int_{-\infty}^{\infty}\frac{{\rm e}^{-z^2}}{z-z_{\rm p}}{\rm d}z={\rm i}s\pi W(s z_p)~,
\end{equation}
where $s={\rm Sgn}[{\rm Im}(z_{\rm p})]$ and Sgn is known as the sign function and its value is $+1$ when the argument is positive and $-1$ when the argument is negative. $W(z)$ is known as Faddeeva function and it is defined as
\begin{equation}
W(z)={\rm e}^{-z^2}{\rm erfc}({-{\rm i}z})~.
\end{equation}
Hence the complex susceptibility is given by
\begin{flalign}
\chi_{\rm D}=&-\frac{{\rm i}{\cal N}d_{21}^2\Omega_{\rm c}^2\sqrt{\pi}}{\hbar\epsilon_0k_{\rm c}^2(k_{\rm c}-k_{\rm p})u^3}\frac{\gamma'}{2\Gamma_1}\times\nonumber\\
&\biggl[-\frac{(z_1-\phi_+)}{[(\beta+\phi_+)^2+\sigma^2](\phi_+-\phi_-)}s_+W(s_+\phi_+)\nonumber\\
&+\frac{(z_1-\phi_-)}{[(\beta+\phi_-)^2+\sigma^2](\phi_+-\phi_-)}s_-W(s_-\phi_-)\nonumber\\
&-\frac{{\rm i}(z_1+\beta+{\rm i}\sigma)}{2\sigma(\beta+\phi_++{\rm i}\sigma)(\beta+\phi_-+{\rm i}\sigma)}W(-\beta-{\rm i}\sigma)\nonumber\\
&+\frac{{\rm i}(z_1+\beta-{\rm i}\sigma)}{2\sigma(\beta+\phi_+-{\rm i}\sigma)(\beta+\phi_--{\rm i}\sigma)}W(-\beta+{\rm i}\sigma)\biggr]~,\nonumber\\
\end{flalign}
where $s_+={\rm Sgn}[{\rm Im}(\phi_+)]$ and $s_-={\rm Sgn}[{\rm Im}(\phi_-)]$

One can use the same approximation as discussed in the article to approximate Eq. (B7) and the approximated expression of Eq. (B7) is given by
\begin{flalign}
&\chi_{\rm D}=-\frac{{\cal N}d_{21}^2\Omega_{\rm c}^2}{\hbar\epsilon_0\sqrt{\pi}k_{\rm c}^2(k_{\rm c}-k_{\rm p})u^3}\frac{\gamma'}{2\Gamma_1}{\rm e}^{-\beta^2}\times\nonumber\\
&\left[\frac{\pi(\beta+z_1)(\phi_--\phi_+)+{\rm i}s_+\pi\sigma(\phi_++\phi_--2z_1)}{\sigma(\phi_+-\phi_-)(\beta+\phi_++{\rm i}s_+\sigma)(\beta+\phi_--{\rm i}s_+\sigma)}\right]~.
\end{flalign}


\section{The exact result of Eq. (13)}
To evaluate the integral in Eq. (13), the integrand can be re-written using partial fractions as
\begin{flalign}
&\frac{{\rm e}^{-z^2}}{(z^2+\sigma^2)(z+\xi+{\rm i}\gamma)}=-\frac{{\rm e}^{-z^2}/2}{\sigma(\sigma-\gamma+{\rm i}\xi)(z+{\rm i}\sigma)}\nonumber\\
&~~~~~~~~~~~~~~~~~~~~~-\frac{{\rm e}^{-z^2}/2}{\sigma(\sigma+\gamma-{\rm i}\xi)(z-{\rm i}\sigma)}\nonumber\\
&~~~~~~~~~~~~~~~~~~~~~+\frac{{\rm e}^{-z^2}}{[\sigma^2+(\xi+{\rm i}\gamma)^2](z+\xi+{\rm i}\gamma)}~.
\end{flalign}
Using Eq. (B5) and re-arranging the expression, the complex susceptibility is given by
\begin{flalign}
\chi_{\rm D}(\Delta_{\rm p})=&-\frac{{\rm i}{\cal N}d_{21}^2\Omega_{\rm c}^2\sqrt{\pi}}{4\hbar\epsilon_0 k_{\rm c}^2(k_{\rm c}-k_{\rm p}u^3)}\times\frac{1}{{\sigma^2+(\xi+{\rm i}\gamma)^2}}\nonumber\\
&\times\left[{\rm e}^{-z_0^2}{\rm erfc}(-{\rm i}z_0)+\frac{{\rm i}z_0}{\sigma}{\rm e}^{\sigma^2}{\rm erfc}(\sigma)\right]~,
\end{flalign}
where $z_0=\xi+{\rm i}\gamma$.


\end{document}